\documentclass[useAMS,usenatbib]{mn2e}

\usepackage{float} 
\usepackage{epsfig}
\usepackage{graphicx}
\usepackage[latin1]{inputenc}
\usepackage{latexsym}
\def\simlt{\ \raise -2.truept\hbox{\rlap{\hbox{$\sim$}}\raise5.truept   %
\hbox{$<$}\ }}
\def\simgt{\ \raise -2.truept\hbox{\rlap{\hbox{$\sim$}}\raise5.truept   %
\hbox{$>$}\ }}                                                          %

\def\be{\begin{equation}}
\def\ee{\end{equation}}
\def\newline{\hfil\break}

\def\la{\mathrel{\hbox{\rlap{\hbox{\lower4pt\hbox{$\sim$}}}\hbox{$<$}}}}
\def\ga{\mathrel{\hbox{\rlap{\hbox{\lower4pt\hbox{$\sim$}}}\hbox{$>$}}}}

\def\MS7{MS 0735.6+7421}

\title[DM annihilation in Coma cluster]{Constraints on dark matter annihilation and turbulent reacceleration set by high-frequency observations of the radio halo in the Coma cluster}
\author[P. Marchegiani et al.]{P. Marchegiani$^{1}$\thanks{E-mail: paolo.marchegiani@inaf.it}, V. Vacca$^{1}$, F. Govoni$^{1}$, M. Murgia$^{1}$, and F. Loi$^{1}$\\ \\
$^{1}$INAF-Osservatorio Astronomico di Cagliari, Via della Scienza 5, I-09047 Selargius (CA), Italy\\
}

\begin{document}

\date{Accepted 2025 August 23. Received 2025 July 17; in original form 2025 April 18.}

\pagerange{\pageref{firstpage}--\pageref{lastpage}} \pubyear{2025}

\maketitle

\label{firstpage}

\begin{abstract}
We study the impact of the recent observation of the radio halo in the Coma galaxy cluster at 6.6 GHz with the Sardinia Radio Telescope on models based on turbulent reacceleration of electrons produced in dark matter annihilation processes. 
Observing at that frequency it is possible to obtain information on the electrons spectrum at energies 
where the effect of turbulent reacceleration becomes sub-dominant with respect to energy losses, and therefore to obtain information 
on the properties of seed electrons. Under the assumption that dark matter particles are neutralino-like particles annihilating at a rate close to the maximum allowed by 
\textit{Fermi}-LAT upper limits in dwarf galaxies, we obtain some constraints on the intensity of the reacceleration and on the value of the neutralino mass. 
In particular, models with mass of the order of 10 GeV are generally 
disfavored, because they produce a high-frequency radio spectrum that can not reproduce the possible flattening observed between 5 and 6.6 GHz; on the other hand, models with mass of the order of 500 GeV, 
in order to reproduce the observed spectrum at frequencies below 100 MHz, 
require a reacceleration phase
longer than $10^9$\,yr, which would require more than one event responsible of the generation of turbulence in the cluster. 
The resulting optimal mass values are in the range 100--200 GeV, with a preference for the quark annihilation channel.
\end{abstract}



\section{Introduction}

Most of the matter content of galaxy clusters is in a ``dark'' form: it does not emit in any band of the electromagnetic spectrum, but reveals its presence through gravitational effects on the velocity distribution of galaxies inside the cluster (e.g. Zwicky 1933) and in gravitational lensing measures (e.g. Clowe et al. 2006). The nature of this dark matter (DM) remains at the moment unknown; considerations based on the comparison between the predictions of primordial nucleosynthesis models and the observed abundance of elements, and on the theory of structures formation, suggest that it should be of a non-baryonic type (see, e.g., Kolb \& Turner 1990).

One of the most promising candidates as DM particle is the neutralino, i.e. the lightest stable particle predicted by super-symmetric theories (see Bertone, Hooper \& Silk 2005 for a general review); this kind of particle can self-annihilate following different channels involving the production of leptons, quarks, or gauge bosons, producing in the final state electrons/positrons and gamma rays. The detection of the associated electromagnetic signal from cosmic structures is in principle an indirect way to probe the presence of DM and possibly determine its nature.

In particular, non-thermal electrons originated in DM annihilation processes inside galaxy clusters can interact with the intra-cluster magnetic field, and produce by synchrotron a diffuse radio emission, which can have similar properties to those of radio halos, i.e. diffuse emissions not related to the single galaxies inside the cluster, having a size comparable to the cluster one and usually a quite regular shape and a low level of polarization (see Feretti et al. 2012, Brunetti \& Jones 2014, van Weeren et al. 2019 for reviews). The spectral shape of the radio emission produced by electrons originated by DM annihilation is related to the properties of the DM particle, as mass and annihilation channel; therefore, if the observed radio emission is dominated by DM-produced electrons, in principle from its spectral shape it could be possible to determine the main properties of DM (Colafrancesco, Profumo \& Ullio 2006).

However, in galaxy clusters other processes can produce or accelerate non-thermal electrons on large scale, like the hadronic interactions between non-thermal protons and thermal nuclei (Blasi \& Colafrancesco 1999), or the acceleration produced by shocks and turbulence originated during clusters merging events (Tribble 1993).
In particular, stochastic particles acceleration due to interactions with electromagnetic waves originated by the cluster turbulence can have a major role in the formation of the radio halos; in this way in fact it is possible to explain why radio halos are usually observed in clusters undergoing or following a major cluster merger (e.g. Cassano et al. 2013), a property that is not understandable if electrons are originated by hadronic interactions or DM annihilation in absence of other processes.

On the other hand, the stochastic acceleration due to turbulence is a low-efficiency process, which can not be effective on the electrons of the thermal gas without over-heating the gas itself (Petrosian \& East 2008). For this reason, it is necessary to assume that this process re-accelerates seed electrons that are already relativistic, like a population of previously accelerated electrons that has aged because of energy losses by accumulating at energies around $\gamma\sim100-200$, for which the electrons lifetime has its maximum (e.g. Brunetti et al. 2001), or the electrons produced in hadronic interactions (e.g. Brunetti et al. 2012).

Another possibility is that the seed electrons are provided by DM annihilation.
In this respect, in a previous paper (Marchegiani 2019) it has been shown that, if the annihilation of neutralino-like particles in the Coma cluster proceeds at a rate close to the upper limits allowed by the gamma ray upper limits established in dwarf galaxies with \textit{Fermi}-LAT (Ackermann et al. 2015), the resulting electrons can produce a radio emission with flux and spectral shape similar to those of the Coma radio halo, when interacting with a magnetic field with properties similar to the observed one (Bonafede et al. 2010), for realistic values of the substructures boosting factor (Pieri et al. 2011), and if a turbulent reacceleration with intensity and duration compatible with what is theoretically expected (e.g. Brunetti \& Lazarian 2007) is active.

However, from the results obtained in Marchegiani (2019) it is also evident that the effect of the turbulent reacceleration is to modify the shape of the produced radio spectrum in a way that is strongly dependent on the intensity and the duration of the reacceleration, making it very difficult to reconstruct the properties of the DM particles starting from the observed shape of the radio spectrum. 

In order to derive better constraints on the DM particles properties, it is important to study the spectrum of the radio halo at high frequencies ($\nu>5$ GHz). In fact, the effect of the reacceleration is expected to be important in the range of electrons energies for which the characteristic time of the reacceleration
 is smaller than the energy loss rate. 
The reacceleration time is given by $\tau_{acc}=p^2/4D_{pp}$, and therefore, if 
the diffusion coefficient in the momentum space has a form of the type $D_{pp}\propto p^2$ (e.g. Brunetti \& Lazarian 2007), 
$\tau_{acc}$ is independent on the electrons energy.
The energy loss rate is instead given by $t_{loss}=\gamma/b(\gamma)$, which, for radiative losses, for which $b(\gamma)\propto \gamma^2$ (e.g. Sarazin 1999), is proportional to $t_{loss}\propto 1/\gamma$.
Therefore, for electrons with energies bigger than the value for which $\tau_{acc}\sim t_{loss}$, the effect of the reacceleration is gradually suppressed with increasing energy, and the spectrum of the electrons at high energies is expected to take the shape of the one of seed electrons even when the reacceleration is active (see, e.g., fig.3 in Brunetti \& Lazarian 2011).

A corresponding behavior is expected to be observable in the radio spectrum: at some frequency, usually at $\nu\simgt 1$ GHz for realistic values of the reacceleration strength and the energy losses (e.g. Brunetti et al. 2001), the effect of the reacceleration is expected to decrease, producing a steepening of the spectrum at higher frequencies, as observed in the Coma radio halo spectrum in the 2--5 GHz range (Thierbach et al. 2003; hereafter T03). But at even higher frequencies the radio spectrum is expected to recover the one produced by seed electrons; for this reason observing this part of the radio spectrum would be very important in order to derive information on the seed electrons and their origin.

Recently, Murgia et al. (2024; hereafter M24) have observed the Coma radio halo at 6.6 GHz with the Sardinia Radio Telescope (SRT), obtaining that the spectrum of the Coma radio halo, even considering the uncertainties due to the error bars and other uncertainties deriving from the procedures used for measuring the flux density at different frequencies (see Appendix A for a discussion), seems to flatten, or at least slow down the steepening, after the frequency of $\sim5$ GHz, compared to the quick steepening observed between 1.4 and 5 GHz. 
In this paper we examine the consequences that this behavior in the radio halo spectrum has on the constraints that can be derived on the properties of the DM particles and of the reacceleration.

In the following, in Sect.\,2 we describe the methods used for the calculations, and in Sect.\,3 we present the results. Finally we discuss the results and present our conclusions in Sect.\,4. In the calculations, we adopt a cosmological model in line with the results obtained with the Planck satellite, with $\Omega_m = 0.308$, $\Omega_{\Lambda} = 0.692$, and $H_0 =67.8$ km s$^{-1}$ Mpc$^{-1}$ (Ade et al. 2016). The luminosity distance of the Coma cluster for this cosmological model at $z=0.023$ is $D_L=104$ Mpc, and 1 arcmin corresponds to 28.9 kpc.

\section{Methods}

We follow the same methods as described in Marchegiani (2019), which we summarize here: we describe the cluster as a spherical symmetric system, with the thermal gas distribution described as indicated by X-ray observations, i.e. with an isothermal beta model with radial profile $n_{th}(r)=n_{th,0}[1+(r/r_c)^2]^{-3\beta/2}$, with $n_{th,0}=3.4\times10^{-3}$ cm$^{-3}$, $\beta=0.75$, and $r_c=0.3$ Mpc, where these parameters have been obtained by adapting to the cosmological model adopted in this paper the results found by Briel, Henry \& Boehringer (1992), and with the magnetic field having a radial profile $B(r)=B_0 n_{th}^{1/2}$, with $B_0=4.7\,\mu$G, as derived by Faraday rotation measures analysis (Bonafede et al. 2010). 

The DM halo is described as a single halo having a Navarro, Frenk \& White (1996) profile, $\rho(r)=\rho_s/[(r/r_s)(1+r/r_s)^2]$, with parameters derived from the scaling laws obtained from fitting to cosmological simulations results once the halo mass is known (Bullock et al. 2001); for the mass of the Coma cluster, $1.24\times10^{15}$ M$_\odot$ (Okabe et al. 2014), we obtain a central density $\rho_s=4.83\times10^3\rho_c$, where $\rho_c$ is the critical density of the Universe, and a scale radius of $r_s=0.545$ Mpc. We don't consider the presence of secondary halos, because it has been found that they can change the shape of the 
spatial distribution of the radio halo surface brightness, making it more similar to the observed one compared to the case of a single DM halo, but a limited impact on the halo flux density, increasing it by a maximum of $10-20\%$, and on the halo spectral shape (Marchegiani \& Colafrancesco 2016).

The annihilation of DM particles is assumed to produce electrons/positrons at a rate
\begin{equation}
Q (E,r) = {\cal B} \langle \sigma v\rangle \frac{dN(E)}{dE}\mathcal{N}_{DM} (r)
\label{source.term}
\end{equation}
(e.g. Colafrancesco et al. 2006), where $dN(E)/dE$ is the spectrum of produced particles, which depends on the DM particle mass $M_{DM}$ and the annihilation channel, and which we calculate using the \textsc{darksusy} package (Gondolo et al. 2004), $\mathcal{N}_{DM} (r) = (\rho (r) )^2/(2 M_{DM}^2)$ is the number of DM particles pairs, $\langle \sigma v\rangle$ is the thermally-averaged annihilation cross section, and ${\cal B}$ is a boosting factor due to the presence inside the DM halo of sub-halos with small mass (Pieri et al 2011). 

In the following, we fix the value of the cross section at the upper limits found using gamma ray observations with \textit{Fermi}-LAT of dwarf galaxies (Ackermann et al. 2015). This means that the results reported in this paper do not violate gamma ray upper limits, but need to be considered as upper limits too, while the real radio spectra due to DM annihilation can be much weaker. 

The value of the boosting factor is not well known; typical values can be estimated to be in the range $30-70$ (e.g. Anderhalden \& Diemand 2013; Sanchez-Conde \& Prada 2014; Ishiyama 2014), but it can be also much higher (e.g. Gao et al. 2012) or smaller (e.g. Storm et al. 2017); in the following we leave ${\cal B}$ as a ``free" parameter, which can be changed in order to recover the observed flux density of the radio halo, and which can be considered acceptable if its value is not too far from typical values estimated in literature.

The spectrum of the electrons produced by DM annihilation evolves in time according to the following equation, written in terms of the electrons momentum $p$:
\begin{eqnarray}
\frac{\partial n(p)}{\partial t} & = & \frac{\partial}{\partial p} \left[ \left(-\frac{2}{p}D_{pp}+\sum_i \left|\frac{dp}{dt}\right|_i \right)n(p) \right. \nonumber\\
 & & \left. + D_{pp}\frac{\partial n(p)}{\partial p}\right] +Q(p)
 \label{eq.diffusion}
\end{eqnarray}
(e.g. Schlickeiser 2002), where the source term $Q$ is given by eq.(\ref{source.term}), the terms $|dp/dt|_i$ are due to the energy losses for different mechanisms, i.e. Coulomb interactions, non-thermal bremsstrahlung, synchrotron, and inverse Compton scattering with the Cosmic Microwave Background (CMB) photons (e.g. Sarazin 1999), and the diffusion coefficient in the momentum space $D_{pp}$ is related to the turbulent acceleration, which we parametrize as $D_{pp}=\chi p^2 /2$, so that the parameter $\chi$ is related to the intensity of the acceleration, and its characteristic time is given by $\tau_{acc}=1/2\chi$.
This form of the diffusion coefficient is predicted in the case of the resonant Transit Time Damping or for stochastic non-resonant compression (Brunetti \& Lazarian 2007),  or in the case of soleinodal turbulence (Brunetti \& Lazarian 2016); in other cases different forms are possible as in the case of turbulent modes on small scales, where $D_{pp}\propto p^w$, with $w$ being the spectral index of the magnetohydrodynamics (MHD) waves (e.g. Fujita et al. 2003; Brunetti et al. 2004), or for magnetic pumping induced by large scale fluctuations (Ley et al. 2023). In this paper we adopt the form shown above because, in addition to being representative of several cases, it allows to describe the acceleration through the parameter $\chi$, which, under this assumption, is independent on the electrons energy and can be quickly linked to the acceleration characteristic time, and can be adjusted phenomenologically comparing the results with the observed data. Under different assumptions, the acceleration characteristic time can be a function of the electrons energy (see, e.g., figs. 12 and 13 in Brunetti et al. 2004); an exploration of these cases can be the subject of future works.

In eq.(\ref{eq.diffusion}) there are some quantities that can depend on the position inside the cluster, as the source term $Q(p)$ (through ${\cal N}_{DM}(r)$), the loss terms $|dp/dt|_i$ (through $n_{th}(r)$ and $B(r)$) , and in principle also $D_{pp}$, because the intensity of the acceleration
can change inside the cluster by effect, e.g., of changes in the magnetic field intensity or of the motion of the galaxies 
(e.g. Brunetti et al. 2001). In the following, since we are interested to the flux density integrated over the volume, we assume that $\chi$ has a constant value inside the cluster. We also neglect the effect of the spatial diffusion, which can have some effect on the surface brightness profile close to the cluster center, but a limited effect on the total flux density (Colafrancesco et al. 2006). In this way, eq.(\ref{eq.diffusion}) can be solved at each $r$ separately, with a significant simplification of calculations.

Eq.(\ref{eq.diffusion}) can therefore be solved by adopting a finite difference scheme (see Donnert \& Brunetti 2014 and references therein for details). In doing this, we assume that the process is proceeding in two phases. 

In the first phase there is not turbulent acceleration ($\chi=0$), corresponding to a phase where the cluster is in a quiet state. During this phase eq.(\ref{eq.diffusion}) can be solved analytically (e.g. Colafrancesco et al. 2006), providing an electrons spectrum that results from the equilibrium between the continuous injection of DM-produced electrons and the effect of energy losses. 
This equilibrium spectrum represents the seed electrons.

In the second phase, following a merging event, the increase of the turbulence level in the cluster gives origin to a diffuse reacceleration of the seed electrons. We assume that the parameter $\chi$ is constant during this phase; this is just an approximation because it is reasonable to think that $\chi$ can have a time evolution depending on the state of the evolution of the electromagnetic waves originated by the turbulence (e.g. Brunetti \& Lazarian 2007). Therefore we can think to the value of $\chi$ we are using as an effective value, resulting from its time evolution and space variations. 
We assume that this reacceleration phase has a duration $T_{acc}$, which we leave as a free parameter, but which is supposed to be no longer than $\sim1$ Gyr, because this is the tipical value for which this kind of process is expected to be effective (e.g. Brunetti \& Lazarian 2007); however, in the presence of more than one event that can generate the cluster turbulence, this phase can also have a longer duration, as resulting from numerical simulations (e.g. Miniati 2015). We note that during this phase the source term $Q(p)$ is not expected to change, because DM particles are not subject to electromagnetic forces, and therefore to the action of turbulent reacceleration. This fact simplifies the calculations, and makes this case different from the case where the seed electrons are secondary electrons of hadronic origin, where the spectrum of relativistic protons is instead subject to the action of reacceleration (Brunetti et al. 2012). 

Finally, the synchrotron spectrum is calculated, and compared with the one of the radio halo, at the end of this second phase, without assuming a further phase with no reacceleration, where the electrons spectrum would be expected to come back to its equilibrium state. This assumption is reasonable for a cluster like Coma, where the level of turbulence, although not so high as in most disturbed clusters, is still sufficiently high to think that turbulent reacceleration is still in place (e.g. Eckert et al. 2017).

With these assumptions, the parameters we can change are therefore the neutralino mass $M_{DM}$, which we assume to be in the range of some tens or hundreds of GeV, i.e. the range that can be effectively constrained by \textit{Fermi}-LAT upper limits, the annihilation channel, which we assume to be $\tau^+\tau^-$, $b\bar b$, or $W^+W^-$, the intensity of the reacceleration, given by the parameter $\chi$, having a value of the order of $10^{-17}-10^{-16}$\,s$^{-1}$ in order to have a characteristic acceleration time comparable with the electrons lifetime due to energy losses for electrons with energies $\gamma\sim10^3-10^4$, the duration of the reacceleration phase $T_{acc}$, and the boosting factor ${\cal B}$, which, once the cross section and the magnetic field are fixed, determines the normalization of the spectrum.

In Table \ref{tab.models} we summarize the properties of the models we use in the next Section.
We note that we compare the resulting radio spectra with observed data
essentially with a visual inspection,  
without choosing the best fit parameters
through a statistical analysis based on $\chi^2$ comparison. 
This is due to two facts: firstly, we explore only a limited set of parameters, without allowing them to vary continuously in some range, because the number of parameters that can be changed and the complexity of calculations would make such an operation too time consuming; therefore the spectra shown in this paper are not results obtained from fitting procedures, but just theoretical estimates obtained for some assumed sets of parameters. Secondly, the scatter of the Coma radio halo data, probably due to the differences in the procedures that have been used to calculate the flux densities and the corresponding error bars in different works (see, e.g., discussions in Brunetti et al. 2013 and M24), would make in any case very difficult to obtain an acceptable fit to the data based on the value of the minimum $\chi^2$.  

\begin{table*}{}
\caption{Models used in this paper, with spectra shown in Figures \ref{spectra.9.60} and \ref{spectra}}
\begin{center}
\begin{tabular}{|*{6}{c}|}
\hline
 $M_{DM}$ & channel & $\langle \sigma v\rangle$ & $\chi$ & $T_{acc}$ & ${\cal B}$ \\
 (GeV) & & (cm$^{3}$ s$^{-1}$) & (s$^{-1}$) & (yr) & \\
\hline 
9 & $\tau^+\tau^-$ & $4\times10^{-27}$ & 0 & - & 250 \\
9 & $\tau^+\tau^-$ & $4\times10^{-27}$ & $2.5\times10^{-17}$ & $5\times10^8$ & 100 \\
9 & $\tau^+\tau^-$ & $4\times10^{-27}$  & $5\times10^{-17}$ & $3\times10^8$ & 60 \\
\hline
60 & $b\bar b$ & $1\times10^{-26}$ & 0 & - & 200 \\
60 & $b\bar b$ & $1\times10^{-26}$ & $2.5\times10^{-17}$ & $8\times10^8$ & 200 \\
60 & $b\bar b$ & $1\times10^{-26}$ & $5\times10^{-17}$ & $5\times10^8$ & 100 \\
\hline
125 & $b\bar b$ & $3\times10^{-26}$ & $5\times10^{-17}$ & $7\times10^8$ & 70 \\
200 & $b\bar b$ & $5\times10^{-26}$ & $5\times10^{-17}$ & $8\times10^8$ & 60 \\
200 & $W^+W^-$ & $5\times10^{-26}$ & $5\times10^{-17}$ & $8\times10^8$ & 80 \\
500 & $W^+W^-$ & $2\times10^{-25}$ & $5\times10^{-17}$ & $1.4\times10^9$ & 30 \\
\hline
 \end{tabular}
 \end{center} 
 \label{tab.models}
\end{table*}

\section{Results}

We examine first two cases that in previous papers (Colafrancesco et al. 2011; Marchegiani \& Colafrancesco 2016) had shown to reproduce quite well the shape of the Coma radio halo spectrum in absence of reacceleration: the case of a neutralino with mass $M_{DM}=9$\,GeV and annihilation channel $\tau^+\tau^-$, and the one with mass $M_{DM}=60$\,GeV and annihilation channel $b \bar b$.

\begin{figure*}
\centering
\begin{tabular}{cc}
\includegraphics[width=\columnwidth]{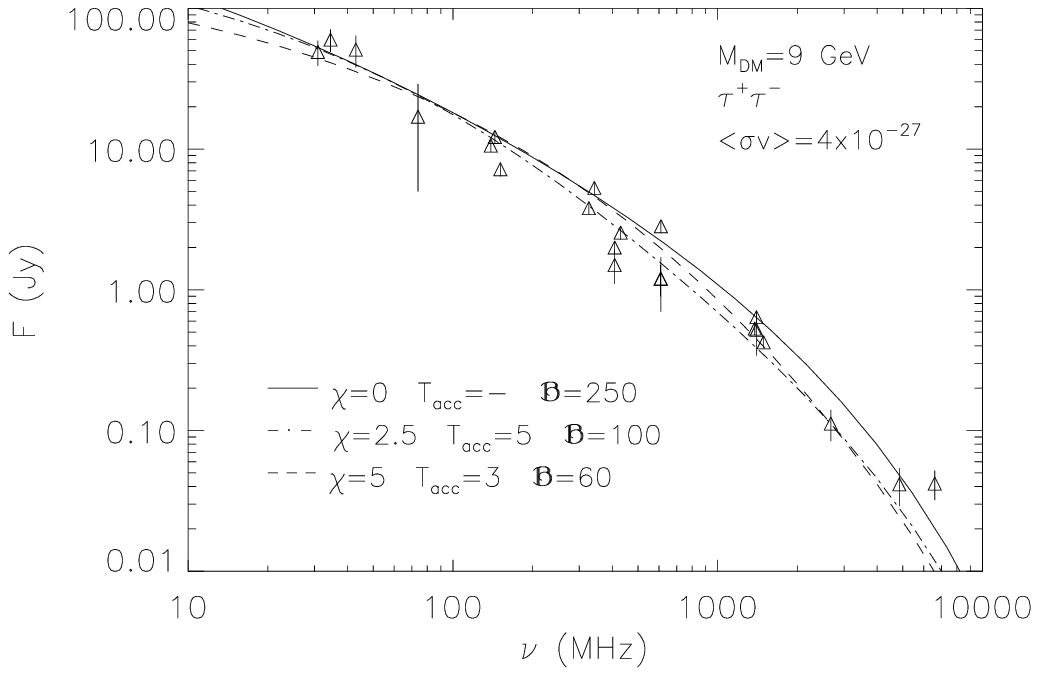} &
\includegraphics[width=\columnwidth]{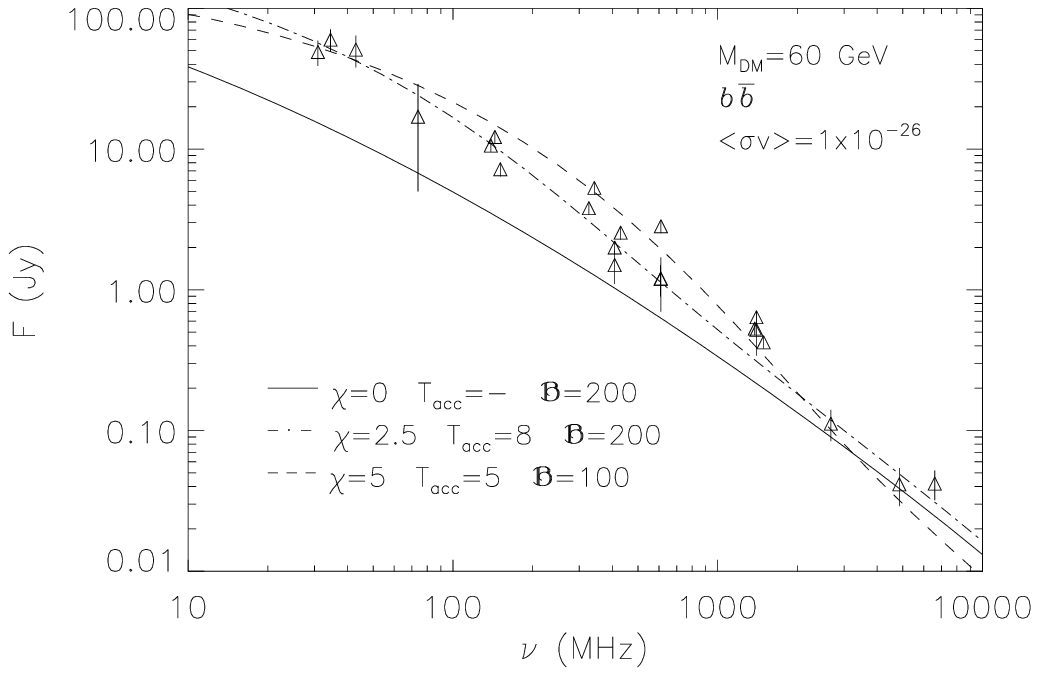} \\
\end{tabular}
\caption{\textit{Left panel}: radio spectra produced for a neutralino with mass $M_{DM}=9$\,GeV and annihilation channel $\tau^+\tau^-$ for no reacceleration and ${\cal B}=250$ (solid line), for reacceleration with $\chi=2.5\times10^{-17}$\,s$^{-1}$, $T_{acc}=5\times10^8$\,yr, and ${\cal B}=100$ (dot-dashed line), and with $\chi=5\times10^{-17}$\,s$^{-1}$, $T_{acc}=3\times10^8$\,yr, and ${\cal B}=60$ (dashed line).
\textit{Right panel}: radio spectra produced for a neutralino with mass $M_{DM}=60$\,GeV and annihilation channel $b \bar b$ for no reacceleration and ${\cal B}=200$ (solid line), for reacceleration with $\chi=2.5\times10^{-17}$\,s$^{-1}$, $T_{acc}=8\times10^8$\,yr, and ${\cal B}=200$ (dot-dashed line), and for $\chi=5\times10^{-17}$\,s$^{-1}$, $T_{acc}=5\times10^8$\,yr, and ${\cal B}=100$ (dashed line).
In the labels inside the plots, $\langle \sigma v\rangle$ is given in units of cm$^{3}$ s$^{-1}$, $\chi$ in units of $10^{-17}$ s$^{-1}$, and $T_{acc}$ in units of $10^8$ yr.
See M24 for a compilation of Coma radio halo data with the corresponding references.}
\label{spectra.9.60}
\end{figure*}

As we can see in left panel of Fig.\,\ref{spectra.9.60}, the model with mass of 9 GeV, which reproduces quite well the shape of the radio halo spectrum until 5 GHz with weak or no reacceleration, can instead be  
challenged by the presence of the point at 6.6 GHz. In fact, the spectrum produced by seed electrons, i.e. the equilibrium spectrum with no reacceleration (solid line in left panel of Fig.\,\ref{spectra.9.60}), shows a steepening that becomes gradually stronger as frequency increases, due to the fact that the spectrum of the electrons tends to zero when their energy approaches $M_{DM}$. The effect of the reacceleration is a further increasing of the spectral curvature, producing a spectrum that can not have a shape after 5 GHz flatter than the equilibrium one. Therefore the presence of the point at 6.6 GHz 
can disfavor models with neutralino mass of the order of 9 GeV. 

With the 60 GeV model the situation is quite different, because the equilibrium spectrum has a high-frequency shape that does not show the fall after 5 GHz as in the 9 GeV case, allowing to predict a flux at 6.6 GHz closer to the observed one. In this case, the shape of the equilibrium spectrum along the whole frequency range (solid line in right panel of Fig.\,\ref{spectra.9.60}) does not match well the observed one, but the presence of the reacceleration can allow to obtain a better accordance with data. In right panel of Fig.\,\ref{spectra.9.60} we show a case with $\chi=2.5\times10^{-17}$\,s$^{-1}$, which provides a spectrum that reproduces quite well the observed one along all the frequency range, with the exception of the points with highest flux densities between 300 and 1400 GHz; reproducing these points would require a higher curvature of the spectrum, which is not achievable with this value of the reacceleration intensity, because in this case the spectrum tends to recover the shape of the equilibrium one at frequencies smaller than 1 GHz. The case with $\chi=5\times10^{-17}$\,s$^{-1}$ instead allows to better reproduce the shape of the spectrum in the central frequency range, but produces a too steep spectrum at higher frequencies which does not reproduce the point at 6.6 GHz. Another problem of the 60 GeV model is that, since the upper limit on the annihilation cross section is quite low, the required substructures boosting factor is quite high, outside the 30--70 range we consider as a conservative one.

\begin{figure*}
\centering
\begin{tabular}{cc}
\includegraphics[width=\columnwidth]{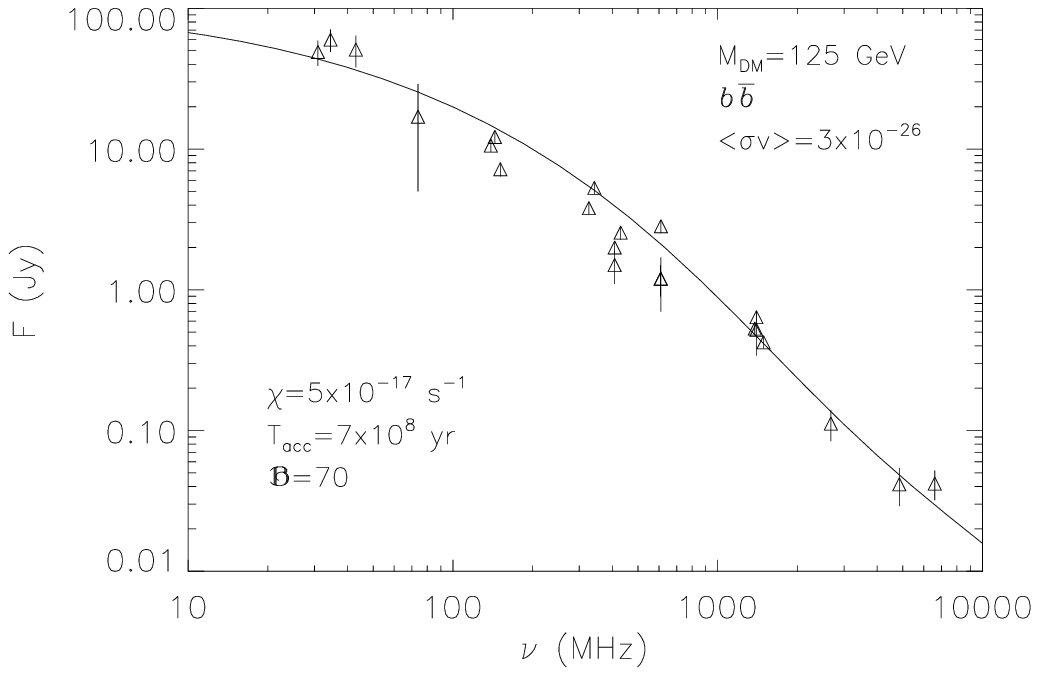}&
\includegraphics[width=\columnwidth]{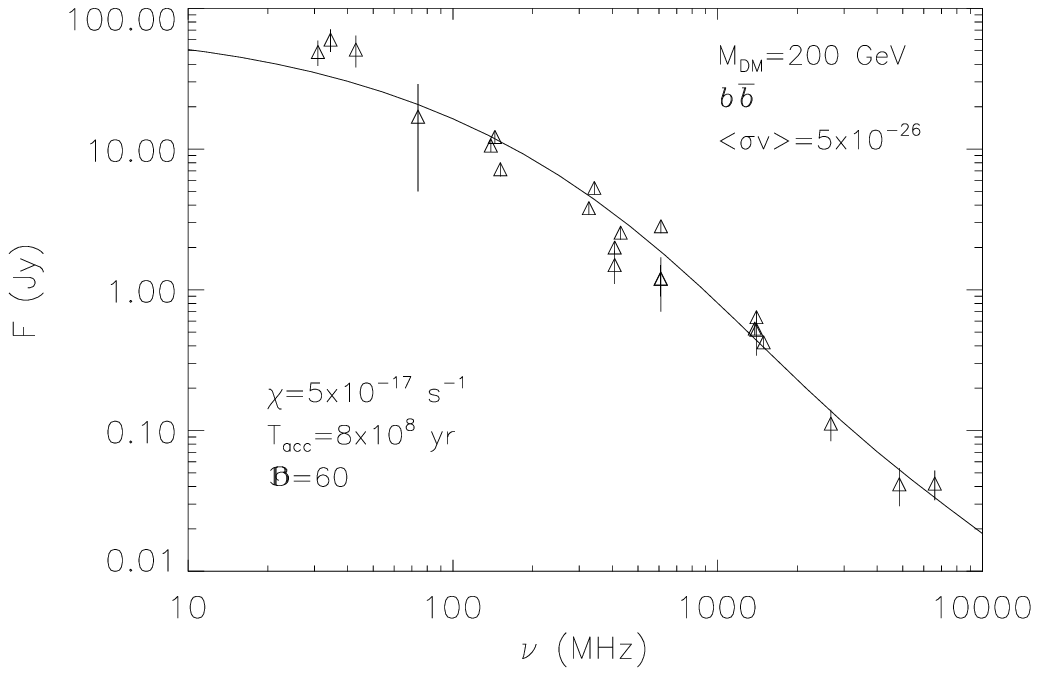} \\
\includegraphics[width=\columnwidth]{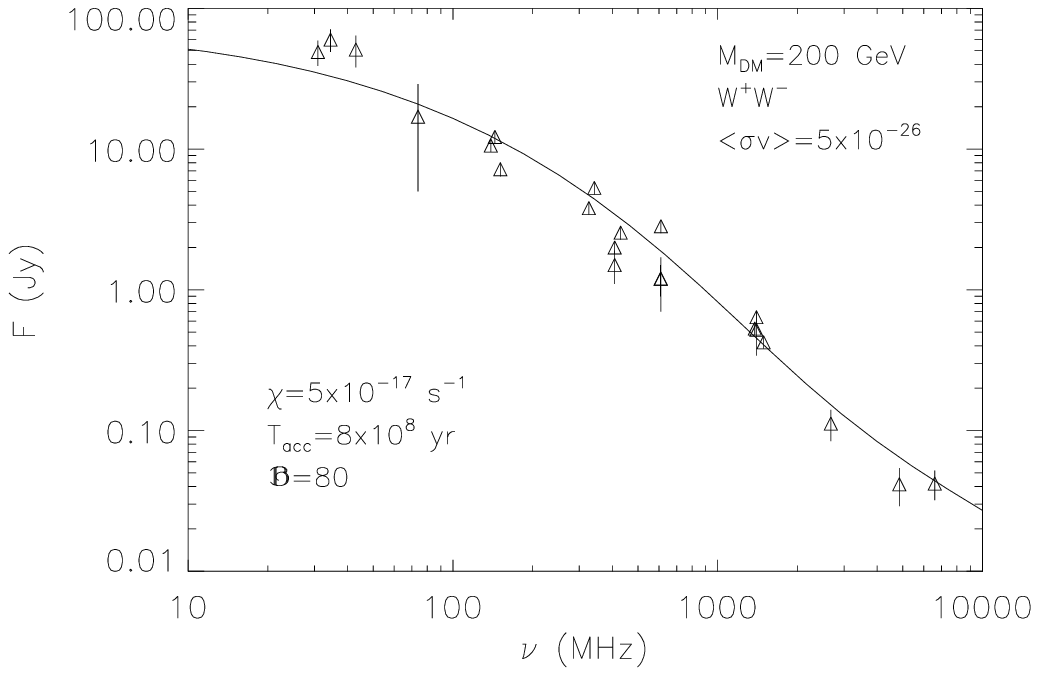} &
\includegraphics[width=\columnwidth]{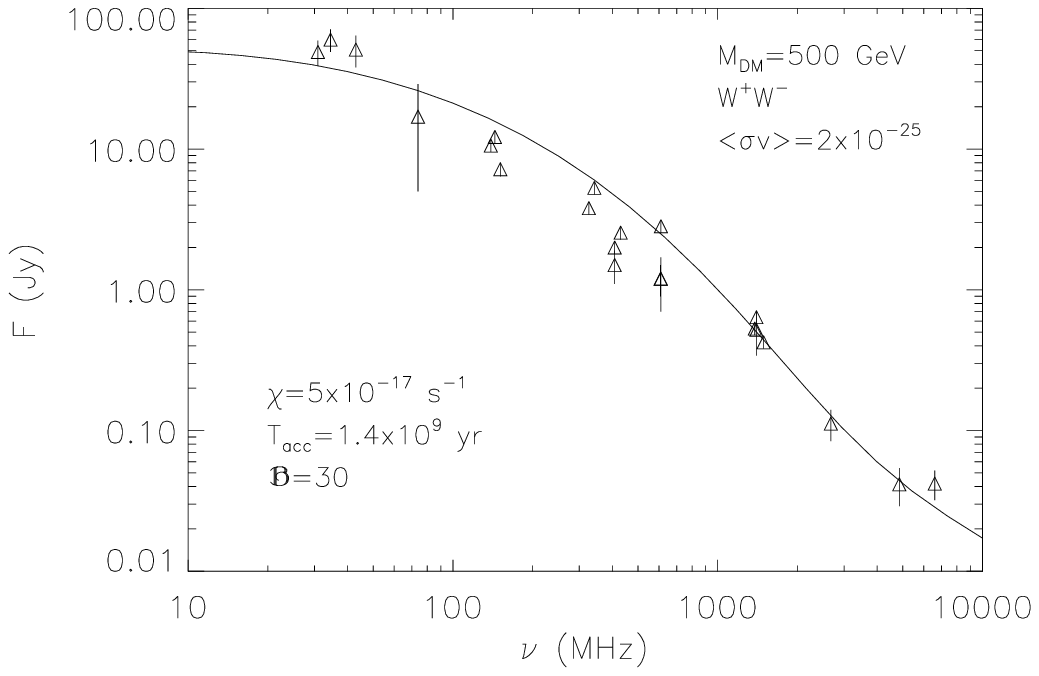} \\
\end{tabular}
\caption{\textit{Top-left panel}: radio spectrum produced for a neutralino with mass $M_{DM}=125$\,GeV and annihilation channel $b \bar b$ for reacceleration with $\chi=5\times10^{-17}$\,s$^{-1}$, $T_{acc}=7\times10^8$\,yr, and ${\cal B}=70$.
\textit{Top-right panel}: radio spectrum produced for a neutralino with mass $M_{DM}=200$\,GeV and annihilation channel $b \bar b$ for reacceleration with $\chi=5\times10^{-17}$\,s$^{-1}$, $T_{acc}=8\times10^8$\,yr, and ${\cal B}=60$.
\textit{Bottom-left panel}: radio spectrum produced for a neutralino with mass $M_{DM}=200$\,GeV and annihilation channel $W^+W^-$ for reacceleration with $\chi=5\times10^{-17}$\,s$^{-1}$, $T_{acc}=8\times10^9$\,yr, and ${\cal B}=80$.
\textit{Bottom-right panel}: radio spectrum produced for a neutralino with mass $M_{DM}=500$\,GeV and annihilation channel $W^+W^-$ for reacceleration with $\chi=5\times10^{-17}$\,s$^{-1}$, $T_{acc}=1.4\times10^9$\,yr, and ${\cal B}=30$.
In the labels inside the plots, $\langle \sigma v\rangle$ is given in units of cm$^{3}$ s$^{-1}$.
See M24 for a compilation of Coma radio halo data with the corresponding references.}
\label{spectra}
\end{figure*}

Therefore we have explored several models with higher mass, and different cases for the annihilation channel. Some significant results are reported in Fig.\,\ref{spectra}, where we show the spectra for the some combinations of parameters we have obtained for the cases with mass 125 GeV and annihilation channel $b \bar b$, mass 200 GeV and channels $b \bar b$ and $W^+W^-$, and mass 500 GeV and channel $W^+W^-$.

By looking at the spectra obtained for these models, we can see that in all these cases the radio halo spectral shape is reproduced quite well at all frequencies, including the 
spectral region after 5 GHz; the 125 GeV model results to be the best one in reproducing the spectral shape also at very low frequencies ($\nu\leq100$ MHz), where other models predict a flatter spectrum than the observed one, and the observed curvature can be recovered only requiring longer acceleration times $T_{acc}$. 
Regarding the annihilation channel, the $W^+W^-$ models produce a flatter spectrum after 5 GHz compared to the $b \bar b$ case; in particular, the 200 GeV model with annihilation channel $W^+W^-$ produces a  
high frequency flattening, which does not allow to reproduce the point at 5 GHz, while the 500 GeV model with annihilation channel $W^+W^-$ produces a better compromise between the point at 5 GHz and the one at 6.6 GHz.

It should also be noted that for higher values of the neutralino mass  
it is generally possible to reproduce the observed flux densities for lower values of the boosting factor. From this point of view, high values of neutralino mass should be preferable. We also found that, in order to reproduce the spectral shape at all frequencies, in particular at $\nu\leq100$\,MHz, higher mass values require longer times for which the acceleration is active, resulting in values of $T_{acc}>1\times10^9$\,yr when $M_{DM}>200$\,GeV. As previously discussed, such values,
at the light of our knowledge about the properties of turbulent reacceleration mechanisms in galaxy clusters, are possible if multiple events generating turbulence take place during such a period.

By putting together all these considerations, we think that models with mass in the range 125--200 GeV, which reproduce the radio halo spectral shape at all frequencies, require reacceleration durations $T_{acc}<1\times10^9$\,yr, which can be reached with a single event producing turbulence, and reasonable boosting factors ${\cal B}\sim70$, can be considered as favorite among those we have considered. In the following we choose the 125 GeV model (top-left panel in Fig.\,\ref{spectra}) as our benchmark model, and use it to study in greater detail how the spectrum changes for different values of the acceleration intensity $\chi$ and duration $T_{acc}$.

\begin{figure}
\centering
\includegraphics[width=\columnwidth]{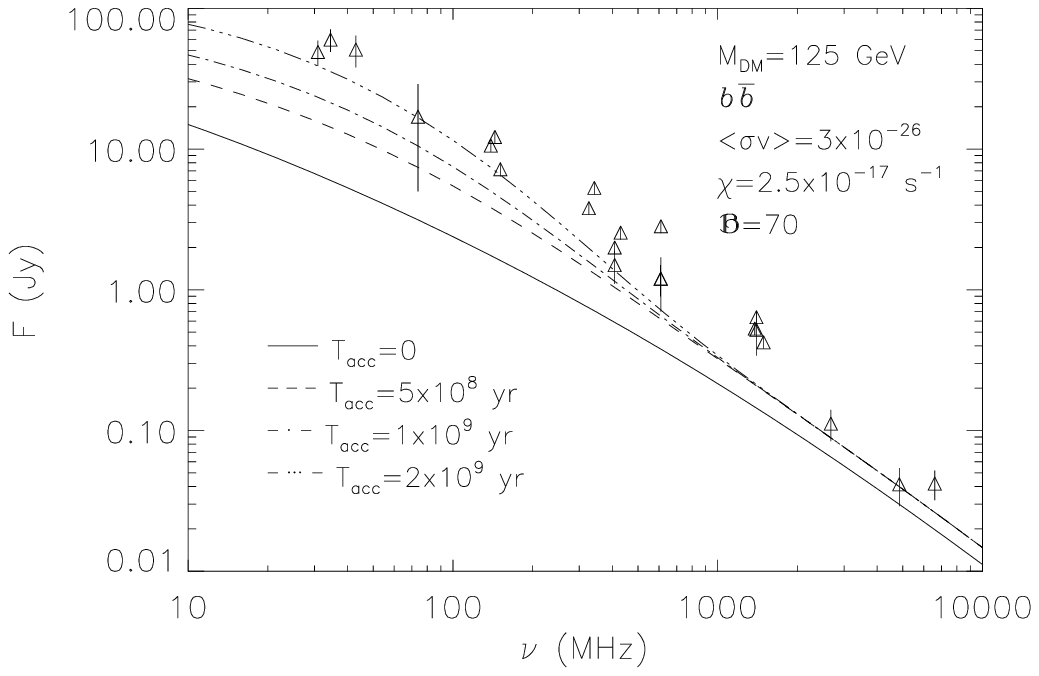} \\
\includegraphics[width=\columnwidth]{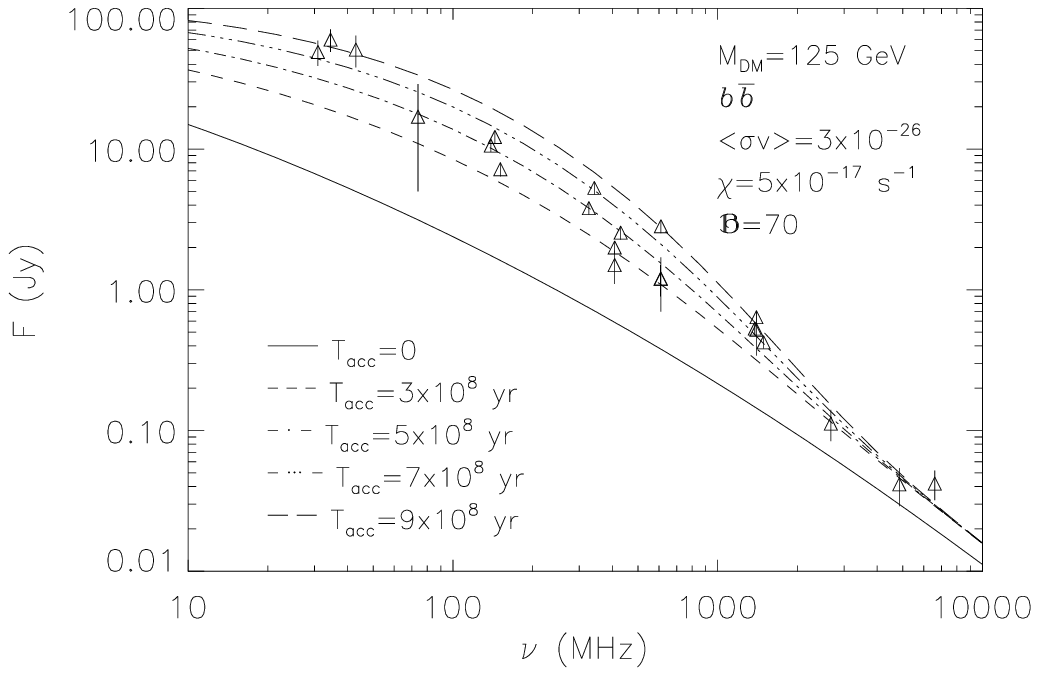} \\
\includegraphics[width=\columnwidth]{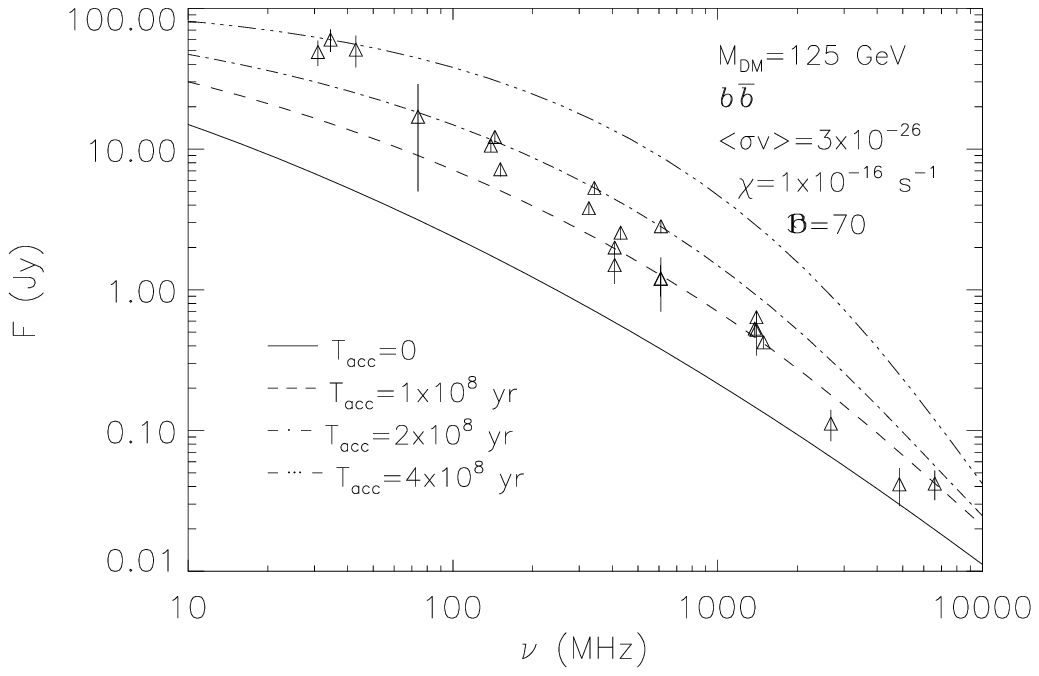}
\caption{\textit{Top panel}: radio spectra produced for a neutralino with mass $M_{DM}=125$\,GeV, annihilation channel $b \bar b$, and ${\cal B}=70$ for no reacceleration (solid line) and for reacceleration with $\chi=2.5\times10^{-17}$\,s$^{-1}$ and $T_{acc}=5\times10^8$\,yr  (dashed line), $T_{acc}=1\times10^9$\,yr (dot-dashed line) and $T_{acc}=2\times10^9$\,yr (three dots-dashed line).
\textit{Middle panel}: as in the top panel, but for reacceleration with $\chi=5\times10^{-17}$\,s$^{-1}$ and $T_{acc}=3\times10^8$\,yr  (dashed line), $T_{acc}=5\times10^8$\,yr (dot-dashed line), $T_{acc}=7\times10^8$\,yr (three dots-dashed line), and $T_{acc}=9\times10^8$\,yr  (long-dashed line).
\textit{Bottom panel}: as in the top panel, but for reacceleration with $\chi=1\times10^{-16}$\,s$^{-1}$ and $T_{acc}=1\times10^8$\,yr  (dashed line), $T_{acc}=2\times10^8$\,yr (dot-dashed line) and $T_{acc}=4\times10^8$\,yr (three dots-dashed line).
In the labels inside the plots, $\langle \sigma v\rangle$ is given in units of cm$^{3}$ s$^{-1}$.
See M24 for a compilation of Coma radio halo data with the corresponding references.}
\label{spectra.125}
\end{figure}

We note that the value of the parameter $\chi$  
resulting as the best one for all the four models we have shown in Fig.\,\ref{spectra} is $\chi=5\times10^{-17}$\,s$^{-1}$. This value results to be well constrained for reproducing at the same time the steepening of the radio halo spectrum between 1.4 and 5 GHz, and the possible flattening between 5 and 6.6 GHz. For better illustrating this point, in Fig.\,\ref{spectra.125} we show, for our benchmark 125 GeV model,  
the effect of decreasing or increasing the value of $\chi$ by a factor of 2, showing also the equilibrium spectrum to better appreciate the modifications produced by the reacceleration for different durations and for a fixed boosting factor.

As we can see, for $\chi=2.5\times10^{-17}$\,s$^{-1}$ the steepening of the spectrum is present at too low frequencies; even increasing the duration of the reacceleration phase up to
$T_{acc}=2\times10^9$\,yr the observed spectral shape can not be reproduced. On the other hand, for $\chi=1\times10^{-16}$\,s$^{-1}$ the spectral curvature due to the effect of the reacceleration is present also at higher frequencies, with a spectrum that continues to steepen also after 5 GHz presenting a too strong curvature after relatively short times. The value of $\chi=5\times10^{-17}$\,s$^{-1}$ is therefore well constrained to well reproduce the observed spectral shape along the whole frequency range, with an acceleration duration around $T_{acc}\sim7\times10^8$ yr producing a spectral curvature similar to the observed one at all the frequencies.

As a final step, we explore the possibility of checking the benchmark model we have chosen in this analysis by calculating the expected emission in the gamma ray band, and by comparing the result with the upper limits found with \textit{Fermi}-LAT in this cluster, and with the expected sensitivity of CTA.

As previously discussed, DM particles are not subject to electromagnetic interactions, and therefore their spectrum is not affected by turbulent reacceleration. As a consequence, the spectrum of gamma rays produced by the decay of neutral pions produced during DM annihilation processes is expected to be the same as the one in absence of reacceleration. Other contributions to the total gamma ray emission can come from the non-thermal electrons when interacting by inverse Compton scattering with the CMB photons and by non-thermal bremsstrahlung with the nuclei of the thermal gas; since these emissions are produced by the electrons, they are instead subject to the effect of the turbulent reacceleration.

\begin{figure}
\centering
\begin{tabular}{cc}
\includegraphics[width=\columnwidth]{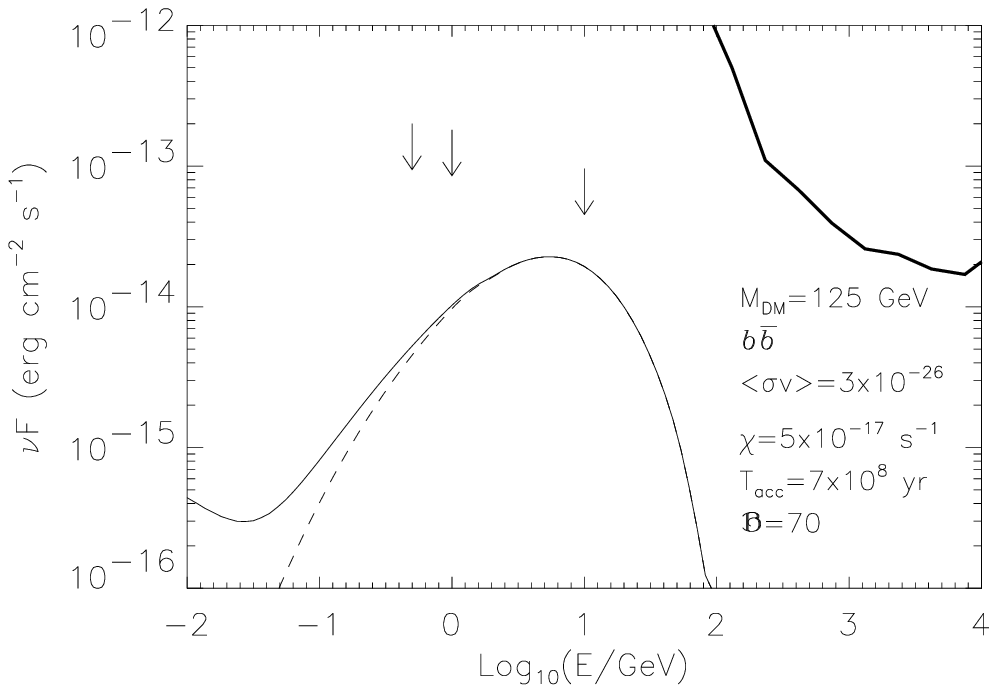}
\end{tabular}
\caption{Gamma ray spectrum produced for a neutralino with mass $M_{DM}=125$\,GeV and annihilation channel $b \bar b$ for reacceleration with $\chi=5\times10^{-17}$\,s$^{-1}$, $T_{acc}=7\times10^8$\,yr, and ${\cal B}=70$. The solid line represents the total emission, while the dashed line represents the emission originated by neutral pions decay only. The arrows represent the \textit{Fermi}-LAT upper limits (Ackermann et al. 2014), and the tick line represents the expected CTA sensitivity for 1000 hours (Funk \& Hinton 2013). In the labels inside the plot, $\langle \sigma v\rangle$ is given in units of cm$^{3}$ s$^{-1}$.}
\label{spectrum.gamma}
\end{figure}

In Fig.\,\ref{spectrum.gamma} we show the results; we can see that the emissions originated by the non-thermal electrons give an appreciable contribution at the total emission only at energies smaller than 1 GeV. The total emission results to be well below the \textit{Fermi}-LAT upper limits: the integrated flux above 100 MeV predicted by this model results to be $F=2.4\times10^{-11}$ cm$^{-2}$ s$^{-1}$, which is a factor $\sim175$ smaller than the upper limit $F_{UL}=4.2\times10^{-9}$ cm$^{-2}$ s$^{-1}$ found in the same band with \textit{Fermi}-LAT (Ackermann et al. 2016). Moreover, since with this value of the neutralino mass, 125 GeV, the gamma ray emission tends to rapidly decrease at energies close to 100 GeV, this emission is not expected to be detectable with Cherenkov telescopes like CTA.

\section{Discussion and conclusions}

The results obtained in this paper show that the addition of the flux density observed with the SRT at 6.6 GHz to the spectrum of the radio halo of the Coma cluster can help to constrain  
the properties of the turbulent reacceleration and of the dark matter particles, if we aim to explain the observed radio emission in terms of annihilation of neutralino particles. This is due to the fact that the spectral flattening possibly observed at $\nu>5$ GHz can be interpreted as produced by electrons with energies higher than the equivalence energy between radiative losses lifetime and reacceleration characteristic time; these electrons are expected to have a spectral shape that tends to take the one of the seed electrons, which is linked to the properties of the dark matter particle, as mass and annihilation channel.

In particular, the value of the reacceleration parameter $\chi$ results to be well constrained around a value of $\chi\sim5\times10^{-17}$\,s$^{-1}$, which corresponds to a reacceleration characteristic time of $\tau_{acc}=1/2\chi\sim1\times10^{16}$\,s, i.e. $\sim3\times10^8$\,yr. We have verified that for values of $\chi$ smaller or higher by a factor of 2 the resulting spectrum has a transition between the steepening and the following flattening at frequencies respectively too small or too high compared to the observed one. We note that this value of the characteristic acceleration time is in line with that can be estimated in post-merger clusters under the assumption of Transit Time Damping interaction between MHD waves and particles (Nishiwaki \& Asano 2025).

Regarding the properties of the dark matter particles, these results show that the models with neutralino mass of the order of 10 GeV, which were favorite in explaining the observed shape of the Coma radio halo spectrum in absence of reacceleration and without the 6.6 GHz point, are instead  
disfavored, because the fall of the spectrum after 5 GHz that is predicted in these models can not be avoided even in presence of reacceleration.

Models with mass of the order of 60 GeV can instead explain the observed possible high-frequency flattening, but 
require values of the substructures boosting factors that, even if not completely excluded, are outside the conservative range of $30-70$.

On the other hand, models with mass of the order of 500 GeV predict an electrons spectrum with a too flat shape along the whole frequency range, unless the reacceleration is active for long times in order to reproduce the observed radio halo spectral curvature at all frequencies, including $\nu\leq100$ MHz. The required duration of the reacceleration phase is higher than $1\times10^9$\,yr, which is expected to be unrealistic for a single event generating turbulence given our knowledge of turbulent reacceleration processes in galaxy clusters (e.g. Brunetti \& Lazarian 2007), but can be obtained in case of several events happening in a comparable time frame (Miniati 2015).

Therefore the favorite neutralino mass range we have found is between 100 and 200 GeV. In particular, the model with mass 125 GeV, annihilation channel $b \bar b$. reacceleration parameter $\chi=5\times10^{-17}$\,s$^{-1}$, reacceleration duration $T_{acc}=7\times10^8$\,yr, and boosting factor ${\cal B}=70$ provides a spectrum with intensity and spectral features that matches quite well the one of the radio halo. 

We need to stress again that these results are obtained assuming that the annihilation cross section is fixed at the upper limits found with \textit{Fermi}-LAT in dwarf galaxies; if the annihilation cross section is noticeably smaller, or if the dark matter has a different nature than neutralino particles, clearly the produced radio emission can not match the observed one, and electrons with different origin need to be invoked to explain the origin of radio halos. Therefore, the results found in this paper do not determine that the seed electrons that, once reaccelerated by turbulent processes, produce radio halos are originated from dark matter annihilation processes, nor that the dark matter particles must have the properties we have derived, but that, if we aim to explain the radio halos in terms of annihilation of neutralino particles, values of the neutralino mass
smaller than 100\,GeV are disfavored, while mass values larger than 200\,GeV require reacceleration durations longer than 1\,Gyr.

Another result is that the predicted gamma ray emission in this scenario is well below the \textit{Fermi}-LAT upper limits found in the Coma cluster, and also below the expected sensitivity of Cherenkov telescopes like CTA. 
As already noted in Marchegiani (2019), this low level of the predicted gamma ray emission makes this scenario different from the one where the seed electrons are produced in hadronic interactions, where instead the predicted gamma ray emission is just below the present \textit{Fermi}-LAT upper limits (e.g. Brunetti et al. 2012); therefore deeper gamma ray observations may help to discriminate between the two scenarios. 

Finally, we discuss about how some of the assumptions made in this paper can be improved in future works.
In this paper we have focused on the flux density integrated over the entire area of the radio halo, considering the possible high-frequency spectral flattening as produced in the whole halo. However, it can not be excluded that the flattening can arise in a specific area of the halo, or can be due to a presence of an additional component emerging at high frequencies (see for example the discussion in Marchegiani \& Colafrancesco 2015 regarding the possibility of different components in the overall spectrum of the radio halo in the Bullet cluster). In this respect, by looking at the spectra in four different sectors of the Coma halo reported in fig.7 in M24, we can see that the spectra in the four sectors look quite similar, with the possible exception of the NW sector, where a flatter spectrum and a higher flux density at 6.6 GHz compared to other sectors are possibly present; therefore it can not be excluded that a component present in this sector (for example related to cosmic ray acceleration due to the shock front detected in the western part of the cluster, see Uchida et al. 2016) has an impact on the overall radio spectrum. In future studies, it will be important to focus also on the spatial properties of the radio halo, considering the distribution of the DM sub-halos inside the cluster as done in Marchegiani \& Colafrancesco (2016), the possibility of a spatial non-uniformity of the reacceleration coefficient (e.g. Brunetti et al. 2001), and studying the spectrum of the halo in different parts of the cluster.

\section*{Acknowledgments}

The Enhancement of the Sardinia Radio Telescope (SRT) for the study of the Universe at high radio frequencies is financially supported by
the National Operative Program (Programma Operativo Nazionale - PON) of the Italian Ministry of University and Research ``Research and
Innovation 2014-2020'', Notice D.D. 424 of 28/02/2018 for the granting of funding aimed at strengthening research infrastructures, in
implementation of the Action II.1 -- Project Proposals PIR01$\_$00010 and CIR01$\_$00010. VV acknowledges support from the Premio per Giovani Ricercatori ``Gianni Tofani'' II edizione, promoted by INAF -- Osservatorio Astrofisico di Arcetri (DD n. 84/2023).
We thank the referee for useful observations and comments that allowed us to improve the quality of the paper and to correct some points.

\section*{Data availability}
No new data were generated or analysed in support of this research.

\appendix

\section{Checking the uncertainties of the possible spectral flattening after 5 GHz}

In this Appendix we discuss about possible uncertainties regarding the real presence of the possible flattening after 5 GHz of the spectrum of the Coma radio halo.

As mentioned in Sect.2, comparing the flux densities obtained with different instruments and calculated adopting different methods can produce the scatter in the overall spectrum, as visible at frequencies $\leq1.4$ GHz, where several measurements are available (see Brunetti et al. 2013 and M24 and references therein); therefore, in principle it is possible that the apparent flattening of the spectrum between 4.85 and 6.6 GHz can be due to analogous reasons. To better check this possibility, we investigated and compared the procedures followed by T03 at 4.85 GHz and M24 at 6.6 GHz. Both these observations were done with single dish telescopes (the 100 m Effelsberg telescope in T03 and the 64 m SRT in M24), so in both cases there should not be the problem of large scale missing fluxes. The procedures were similar, with an integration of the surface brightness inside an area of 35 arcmin of diameter in M24, and inside polygons in T03 inside an area not clearly specified, but which from fig.3b in T03 seems to have a similar extension. In both cases, the subtraction of point sources has been performed by estimating the flux of each source at the appropriate frequency from a spectral fit to the available data at other frequencies; in some cases, especially for weak sources, this resulted in an extrapolation to high frequencies of the spectrum estimated at lower frequencies, which can result in an overestimation of the contribution of point sources. However, the similarity in the procedures makes the estimated point sources contribution similar in the two papers; by extrapolating the total contribution found in T03 at 4.85 GHz, 230 mJy, to the frequency of 6.6 GHz using spectral indices in the range 0.8--1.5 as typical of these sources, it is possible to obtain values in the range 145--180 mJy, i.e. smaller or similar to the value estimated in M24, 181 mJy. Therefore, it seems there is not an overestimation of the point sources contribution in T03 compared to M24 that could have produced the apparent flattening in the spectrum.

Another possible cause that can alter the high-frequency spectral shape is the correction for the Sunyaev-Zel'dovich effect (SZE); the necessity of correcting for this effect comes from the fact that the CMB in the direction of the cluster has a surface brightness which, in the Rayleigh-Jeans part of the spectrum, is weaker than the CMB outside the cluster because of the inverse Compton scattering with the cluster thermal gas; therefore, once the baseline level of the radio image is subtracted, this can result in an underestimation of the halo surface brightness at high frequencies (En\ss lin 2002). The data at frequencies $\geq2.6$ GHz used in this paper are all corrected for the SZE following the procedure described in M24\footnote{We note that the corrections for the SZE applied by M24 at 2.675 and 4.85 GHz were affected by a material error that underestimated the correction, and an erratum to that paper is in preparation (Murgia et al., private communication); in this paper we are using the data with the fixed corrections for the SZE}; the possible sources of uncertainties in applying this correction are given by the estimated value of the average Compton parameter in the halo area (assumed to be the same in T03 and M24), whose uncertainty is taken into account through the error bar of the corrected value of the surface brightness (eq.5 in M24), and by the extension of the halo area in the derivation of the correction in terms of flux density (eq.6 in M24). Another possible source of uncertainty in the correction for the SZE is given by the procedure of the baseline subtraction in the radio image, which, if performed at a distance from the cluster where a residual SZE is still present (which for Coma extends up to a distance comparable to $R_{500}$, i.e. $\sim47$ arcmin, see Ade et al. 2013), can produce on overestimation of the halo flux density once the SZE correction is applied (Brunetti et al. 2013). Comparing the maps of the radio halo at 4.85 GHz and 6.6 GHz (fig.3a in T03 and fig.1 in M24), we can see that the first map extends in a region with size about 1.5 degrees with the cluster at the center, covering therefore a region where the SZE is almost entirely present, while the second one extends for 2 degrees with the cluster close to the eastern side, allowing therefore to cover a region on the western side where the SZE is possibly not present; therefore, it is possible that the corrections applied by M24 to both the maps are overestimating the halo flux density at 4.85 GHz, but not at 6.6 GHz. In conclusion, the main source of possible uncertainties seems to be the extension of the halo area, which impacts on both the integration of the surface brightness and the correction for the SZE.

\bsp

\label{lastpage}

\end{document}